\documentclass{svproc}
\usepackage{url}

\usepackage{graphicx}

\newcount\eqnumber
\def\neweq{{\rm{(\the\eqnumber)}}\global\advance\eqnumber by 1}
\def\eqdef#1{\eqno\xdef#1{\the\eqnumber}\neweq}
\def\newaeq{{\rm{(\the\eqnumber a)}}\global\advance\eqnumber by 1}
\def\eqdaf#1{\eqno\xdef#1{\the\eqnumber}\newaeq}
\def\eqdisp#1{\xdef#1{\the\eqnumber}\neweq}
\def\eqdasp#1{\xdef#1{\the\eqnumber}\newaeq}

\newcount\refnumber
\def\newref{{\the\refnumber}\global\advance\refnumber by 1}
\def\refdef#1{{\xdef#1{\the\refnumber}}\newref}

\newcount\fignumber
\def\newfig{{\the\fignumber}\global\advance\fignumber by 1}
\def\figdef#1{{\xdef#1{\the\fignumber}}\newfig}

\def\smallskip{\vskip 3pt}
\def\medskip{\vskip 6pt}
\def\bigskip{\vskip 12pt}

\begin{document}
\mainmatter              
\title{Singularity interactions for lattice equations: introducing the taishi}
\titlerunning{Lattice Singularities}  
%
%
\author{Ralph Willox\inst{1}, Basil  Grammaticos\inst{2},  Alfred Ramani\inst{2} \and Thamizharasi Tamizhmani\inst{3}}
\authorrunning{Willox et al.} 
%
\tocauthor{Ralph Willox, Basil  Grammaticos, Alfred Ramani and Thamizharasi Tamizhmani}
\institute{Graduate School of Mathematical Sciences, the University of Tokyo, 3-8-1 Komaba, Meguro-ku, 153-8914 Tokyo, Japan
\and Universit\'e Paris-Saclay and Universit\'e Paris-Cit\'e, CNRS/IN2P3, IJCLab, 91405 Orsay, France\ \email{(bgrammat@ijclab.in2p3.fr)}
\and
SAS, Vellore Institute of Technology, Vellore - 632014, Tamil Nadu, India
}

\maketitle              

\begin{abstract}
We study the singularities of some selected integrable lattice equations and show they all admit three types of singularities: one of finite and two of infinite extent. In particular, we show that all the equations we study possess a recently discovered, strip-like, singularity which is known under the moniker of ``taishi''. We study in detail the interaction of these taishi with the remaining two types of singularities and show that the rich behaviour first obtained in the case of the Korteweg-deVries equation is also present for other equations. Moreover, we find that in all cases this behaviour can be encoded in terms of simple symbolic dynamics which are just the dynamics governing a Box \& Ball cellular automaton. 
\end{abstract}
\section{ Introduction}
The integrability properties of a dynamical system are intimately related to its singularity structure [\refdef\bountis]. Already, at the beginning of the previous century,  the work of Painlev\'e and his school pioneered the derivation of new integrable differential systems based on the study of their singularities. In the domain of discrete integrable systems the situation is even more interesting: it was by studying the singularity structure of a known integrable equation that the singularity confinement property was discovered [\refdef\sincon]. As explained in several works of the authors, singularity confinement is a property shared by discrete systems integrable through spectral methods and can, through the procedure known as ``deautonomisation'' [\refdef\desoto], be used as a discrete integrability detector.

While studying the lattice Korteweg-de Vries (d-KdV)  equation [\refdef\hirota],
$$x_{n+1,m+1}-x_{n,m}={1\over x_{n+1,m}}-{1\over x_{n,m+1}},\eqdef\zena$$
it was remarked that when (due to the initial conditions) $x$ happens to take the value 0, the singularity that ensues does not propagate ad infinitum. Figure \figdef\one\ below shows that the singularity is confined to just one lattice square.
\medskip
\centerline{\includegraphics[width=5 cm,keepaspectratio]{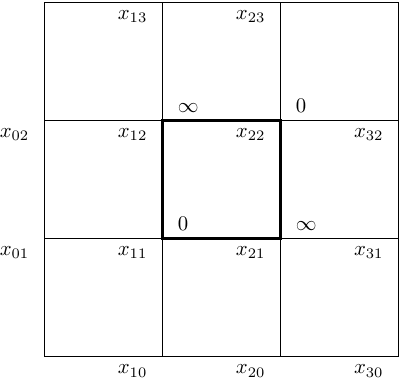}}
\smallskip{\bf Figure \one}. {\sl The elementary singularity of the discrete KdV equation (\zena). The $n$ direction on the lattice is taken as the horizontal one and $m$ as the vertical one in this and all subsequent figures}
\smallskip
Evolving past the singularity in the northeast direction leads to finite values. In fact, there exist, what in numerical analysis are called ``singular rules'' [\refdef\wynn], relating the values of the dependent variable on either side of the singularity:
$$x_{2,3}-x_{1,0}={1\over x_{2,0}}-{1\over x_{1,3}},\eqdaf\zdyo$$
$$x_{3,2}-x_{0,1}={1\over x_{3,1}}-{1\over x_{0,2}}.\eqno(\zdyo{\rm b})$$
The singularity depicted in Figure \one\ is the simplest confined singularity for discrete KdV. Infinitely many more can arise, created by values of 0 adjacent on an oblique line oriented in the northwest-southeast direction. However, these are not the only singularities of the d-KdV. Three more types of singularities do exist. The first two consist of infinite lines of infinities oriented in the southwest to northeast direction or of two such lines with a line of zeros in between them (see [\refdef\kdv] for more details).  But the most surprising singularity is the third one. Rewriting (\zena) as 
$${x_{n+1,m+1}x_{n+1,m}-1\over x_{n+1,m+1}x_{n+1,m}}={x_{n,m}x_{n,m+1}-1\over x_{n,m}x_{n,m+1}},\eqdef\ztri$$
we remark that if $x_{n,m+1}=1/x_{n,m}$ then this relation propagates in the $n$ direction, leading to an infinite strip of values with product 1 over two lines adjacent in the vertical direction. This singularity was named taishi (which derives from the Japanese naming of elementary particles, always ending in -shi, and the character for `strip').  Taishi interact with the other singularities of d-KdV, leading to very interesting dynamics. An example of the interaction of a taishi with a line of infinities is given in Figure \figdef\two.
\medskip
\centerline{\includegraphics[width=6 cm,keepaspectratio]{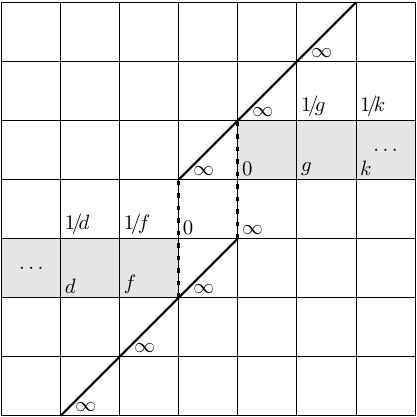}}
\smallskip{\bf Figure \two}. {\sl Interaction of the taishi with a line of infinities}
\smallskip
As explained in [\kdv], in order to have access to the full, unfettered, richness of possible interactions we must first introduce the notion of `weights' for the singularities. In the above, when we write 0 or infinity this must be understood as the limit of $\epsilon$  or $1/\epsilon$ for $\epsilon\to0$ . However, one can imagine zeros or infinities with higher multiplicities, i.e. stemming from powers of $\epsilon$ or $1/\epsilon$. Similarly we can imagine that the values that appear in the taishi are such that their product is equal to $1+{\cal O}(\epsilon^p)$, for some positive integer $p$. In what follows we shall refer to these exponents as ‘weights’.

A detailed study [\kdv] of the taishi interactions with an oblique line of weight $q=1$ (meaning that the values on the line diverge as $1/\epsilon$), which may look complicated on the surface, revealed an underlying process best formulated in the form of symbolic dynamics as follows. Starting sufficiently low,

$\bullet$ move upwards up to the first (i.e. lowest) horizontal strip in a taishi with non-zero weight and subtract 1 from its weight and add 1 to the weight of the strip just above it,

$\bullet$ then move to the next horizontal strip with non-zero weight above those two strips, subtract 1 from its weight and add 1 to the weight of the strip just above it,

$\bullet$ repeat the same procedure until there are no more taishi with strips with non-zero weight left.

The discrete KdV equation is somewhat special since it requires that this procedure be repeated twice in order to lead to the final result. It is as if the line of infinities had an effective weight of 2. However the situation becomes simpler when one studies a similar phenomenon in the case of a discrete modified KdV (mKdV) equation proposed by Levi and Yamilov [\refdef\levi]:
$$(v_{n,m}v_{n,m+1}-1)(k v_{n+1,m+1}+v_{n+1,m}/k)=(v_{n+1,m}v_{n+1,m+1}-1)(k v_{n,m}+v_{n,m+1}/k),\eqdef\ztes$$
where $k$ is a non-zero constant such that $k^4\neq1$, lest the equation become degenerate. A taishi, obviously, exists here as well, corresponding to $v_{m,n}v_{m+1,n}=1$. The line of singularities that plays the role of the line of infinities in this case, is an infinite line with values $1/k$, and when we say that the singularities on this line have weight $q$, we are assuming that they have values $1/k+{\cal O}(\epsilon^q)$. In this case the symbolic dynamics of the taishi are simpler since the effective weight of the oblique line is indeed equal to its ``formal'' weight, as defined above.

The most interesting result is that the symbolic dynamics of the taishi interactions can be formulated in terms of Box\&Ball  [\refdef\takasa] systems. It goes like this.

We represent the weights of all the taishi that will interact with a line of infinities of weight 1, as an infinite column vector $W$ (i.e., a column vector parallel to the $m$ axis of the lattice on which the interaction takes place). The entries in this vector are denoted as $W_0^m$ and encode the weight of the taishi in the strip at vertical position $m$ (where that weight is taken to be zero in the absence of a taishi) and we interpret this vector as depicting an infinite column of boxes, each containing $W_0^m$ balls. We then picture a carrier $V$ that can carry at most one ball, running up the column of boxes $W$ from $m=-\infty$ to $m=+\infty$. We denote the contents of the carrier when it arrives at position $m$ in the column as $V_0^m$. The carrier is empty at $m=-\infty$ but, moving upwards in $m$, at its first encounter with a box $W_0^m$ that does contain a ball, it picks up one ball from $W_0^m$ and immediately deposits it in the box $W_0^{m+1}$ just above it. The (now empty) carrier then moves further upwards, until it reaches the next non-empty box, picking up a ball from it and depositing the ball in the box immediately above that box, and so on. The changes induced in the carrier and box contents by the carrier moving upwards along the boxes can be summarized as
$$V_0^{m+1} = \min[W_0^m,1-V_0^m]\quad {\rm and}\quad W_1^m=W_0^m+(V_0^m-V_0^{m+1}),\eqdef\bbs$$
where $W_1^m$ denotes the contents of the box at position $m$ after the carrier has passed. This is nothing but the update rule obtained from the general Box\&Ball system due to Takahashi and Matsukidaira [\refdef\takama] for the case of boxes with infinite capacity and a carrier with capacity 1. It is readily seen that this system is dual to the famous Takahashi-Satsuma Box\&Ball system [\takasa] upon interchanging carrier and boxes, and the roles of the coordinates depicting space and time.

We shall show one last figure, of taishi interactions in the case of KdV, where the formal weight of the oblique line is 1 and we have multiplied the weights of the taishis by 2 (to be congruent with the effective weight of the line). In Figure \figdef\three\ we show how a taishi of weight 2, during successive interactions with lines of infinities, catches up with a ``heavier'' taishi of weight 4 and overpasses it. If one follows the process in the vertical direction one immediately sees the solitonic behaviour of a Box\&Ball system.

Given the remarkable properties of this newly-discovered singularity, the natural question is whether its existence is limited to systems where the quantity $x_{n,m}x_{n,m+1}-1$ appears naturally. The first step towards answering this is to analyse equations related to the two mentioned above and look for singularities similar to the taishi. It turns out that strip-like singularities do exist, as well as oblique infinite lines of singularities, and that their mutual interaction follows the symbolic dynamics presented in this introduction. 
\medskip
\centerline{\includegraphics[width=6 cm,keepaspectratio]{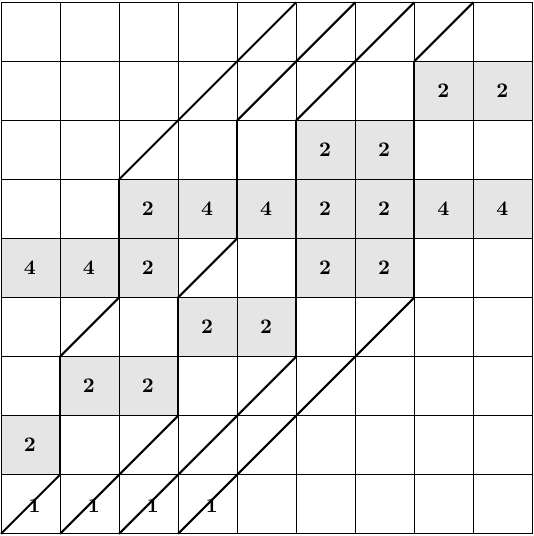}}
\smallskip{\bf Figure \three}. {\sl Interaction of two taishis with a succession of lines of infinities. In order not to overburden the drawing we have given only the weights of the taishis at each point, the weights of the infinities on the oblique lines being 1}
\smallskip

In what follows we shall study three lattice equations: a KdV variant proposed by Levi and Yamilov [\levi], the equation known as the (potential) modified KdV [\refdef\capel] and a sine-Gordon variant [\refdef\heredero] proposed by Heredero and collaborators.
\section{The Levi-Yamilov KdV equation}
Levi and Yamilov proposed in [\levi] two partial difference equations and showed their integrability. These equations were analysed by some of the present authors [\refdef\notso] and shown to be discrete analogues of KdV and its modified counterpart. As mentioned in the Introduction, the latter equation was studied in [\refdef\ourmkdv] where it was shown that it possessed strip-like singularities which were indeed taishis. Curiously the KdV analogue of Levi-Yamilov has never been studied from the taishi perspective. This situation will be remedied in this section.

The Levi-Yamilov KdV has the form
$$(u_{n+1,m}+1)(u_{n,m}-1)=(u_{n+1,m+1}-1)(u_{n,m+1}+1).\eqdef\zoct$$
It is related to the ``standard'' KdV (\zena) by a simple Miura transformation, but only half of it was given in [\refdef\scimi]. We complement it here. The complete Miura is
$$x_{n+1,m}x_{n,m}={1+u_{n,m}\over 1-u_{n,m}},\eqdaf\zenn$$
$$x_{n,m+1}x_{n,m}={1+u_{n,m}\over 1-u_{n,m+1}}.\eqno(\zenn{\rm b})$$
Using (\zenn{\rm a}) we can express $u$ in terms of $x$ as
$$u_{n,m}={x_{n+1,m}x_{n,m}-1\over x_{n+1,m}x_{n,m}+1}.\eqdef\dzer$$
The condition for the existence of a taishi in the case of KdV is $x_{n,m+1}x_{n,m}=1$ and it is straightforward to verify that this entails
$$u_{n,m+1}+u_{n,m}=0\eqdef\dena$$
for the Levi-Yamilov KdV. Moreover, using  (\zoct), one can verify that if $u_{n,m+1}+u_{n,m}=0$ then $u_{n+1,m+1}+u_{n+1,m}=0$ and the condition propagates along a horizontal strip. It is easy to derive the ``singular rule'' relating the values of the dependent variable on either side of the strip singularity:
$$(u_{n+1,m-1}+1)(u_{n,-1}-1)=(u_{n+1,m+2}-1)(u_{n,m+2}+1).\eqdef\ddyo$$
The elementary singularity of (\zoct) was obtained in [\notso]. When $u_{n,m}$ happens to take the value $-1$, a singularity appears which is confined in two steps, as shown in Figure \figdef\four, below. However this is not the only possibility. If more than one adjacent values of $-1$ exist the singularity pattern becomes more complicated but the singularity is eventually confined.

\medskip
\centerline{\includegraphics[width=8 cm,keepaspectratio]{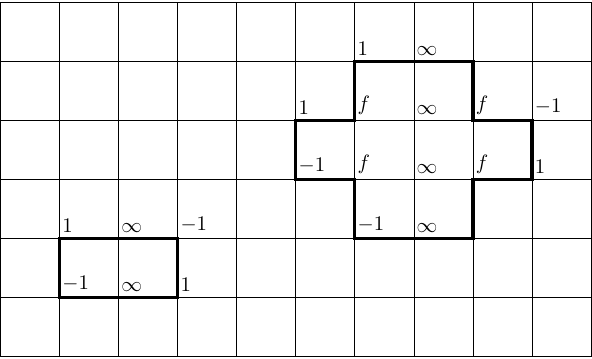}}
\smallskip{\bf Figure \four}. {\sl Confined singularities of (\zoct), where $f$ is a finite value obtained from (\zoct) when $u_{n+1,m}$ and $u_{n,m+1}$ take the value $-1$}
\smallskip
Equation (\zoct) has another singularity of infinite extent, besides the taishi. It is easy to show that if $u_{n,m}=1$ then $u_{n+1,m+1}=1$ and thus one has an oblique line running in the south-west to north-east direction with all values equal to 1. It is precisely with this line that the taishi can interact, leading to interesting behaviour. (To be fair, the taishi can also interact with the confined singularities. Depending on whether the interaction is head-on or off-centre, the confined singularity pattern domain remains as is or is complemented by a pattern that makes it symmetric around the horizontal strip. The taishi, on the other hand, crosses the singularity without being deflected. In [\kdv] we gave examples of such singularities in the case of KdV (\zena). Similar patterns can be produced for the Levi-Yamilov lattice). 

As in the case of KdV presented in the Introduction, we are going to assign weights both to the taishi and the infinite singular line. We will assume that on the horizontal strip we have $u_{n,m+1}+u_{n,m}={\cal O}(\epsilon^p)$ and that on the oblique line  $u_{n,m}=1+{\cal O}(\epsilon^q)$. We should point out here that the end result of an interaction of a taishi with an oblique line of weight $q$ is the same as $q$ interactions with an oblique line of weight 1. Note that in the present case the effective weight of the oblique line is equal to its formal weight, just as in the case of equation (\ztes).
The simplest interaction occurs when a taishi of weight 1 interacts with an oblique line also with weight 1. The result is presented in Figure \figdef\five.
\medskip
\centerline{\includegraphics[width=5 cm,keepaspectratio]{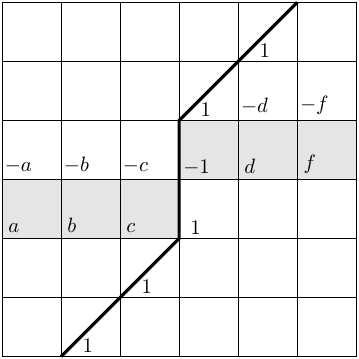}}
\smallskip{\bf Figure \five}. {\sl An interaction of a taishi of weight 1 with an infinite line also of weight 1. Note that in the vertical segment, a value $-1$ appears during the interaction}
\smallskip
The phenomena of ``fusion'' and ``fission'' of the taishi observed in [\kdv] and [\ourmkdv] are also present here. In Figure \figdef\six\ we show a situation where two adjacent taishi of weight one interact with an infinite singular line and collapse into a single one with weight two. 
\medskip
\centerline{\includegraphics[width=5 cm,keepaspectratio]{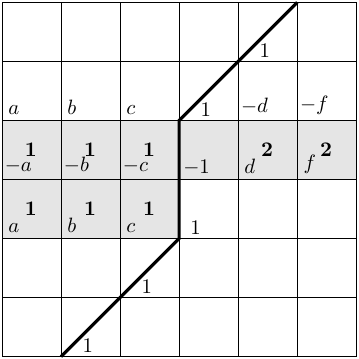}}
\smallskip{\bf Figure \six}. {\sl Interaction of two adjacent taishi of weight 1 with an infinite line also of weight 1, leading to a single emerging taishi of weight 2}
\smallskip
Conversely, in Figure \figdef\seven\ we present an interaction of a weight-2 taishi with an oblique line of weight 1. The impinging taishi is now split into two strips of weight 1.
\medskip
\centerline{\includegraphics[width=5 cm,keepaspectratio]{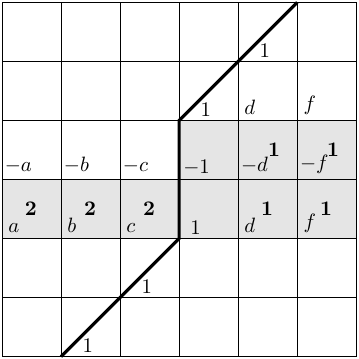}}
\smallskip{\bf Figure \seven}. {\sl  Interaction of a taishi of weight 2 with an infinite line also of weight 1, leading into two emerging taishi of weight 1}
\smallskip
While these interactions may appear complicated they become quite simple to understand in terms of the symbolic dynamics.

Using the prescription presented in the Introduction, the dynamics corresponding to figures \five, \six\ and \seven\ can be simply represented by the diagrams below:
$$\matrix{ 0\cr 0\cr 0\cr \bf1\cr 0\cr 0}\Longrightarrow\matrix{0\cr0\cr \bf1\cr0\cr0\cr0}\hskip2.5cm\matrix{ 0\cr 0\cr\bf1\cr \bf1\cr 0\cr 0}\Longrightarrow\matrix{0\cr 0\cr\bf2\cr0\cr0\cr0}\hskip2.5cm\matrix{ 0\cr 0\cr 0\cr \bf2\cr 0\cr 0}\Longrightarrow\matrix{0\cr0\cr \bf1\cr \bf1\cr0\cr0}$$
where numbers in boldface correspond to the weights in the taishi before interaction (on the left) and after it has left the interaction region (on the right).

More complicated situations can be (and have been) analysed and the upshot is that the singularities of the Levi-Yamilov KdV equation follow the same pattern, when it comes to the taishi interaction, as for the equations studied in [\kdv] and [\ourmkdv].
\section{The (potential) modified KdV equation}
The next equation we are going to study is the potential modified KdV lattice equation. It has been proposed by Capel and collaborators [\capel] and, independently, by Hirota [\refdef\hirom]. It has the form 
$$w_{n+1,m+1}=w_{n,m}{\mu w_{n,m+1}+w_{n+1,m}\over w_{n,m+1}+\mu w_{n+1,m}},\eqdef\dtri$$
where $\mu$ is a parameter. In [\scimi] we have given the Miura transformation of (\dtri) to the Levi-Yamilov modified-KdV (\ztes)
$$v_{n,m}=k{w_{n,m}-w_{n-1,m-1}\over w_{n,m}+w_{n-1,m-1}}={1\over k}{w_{n+1,m}-w_{n,m+1}\over w_{n+1,m}+w_{n,m+1}},\eqdef\dtes$$ 
in terms of a parameter $k$ related to $\mu$ through $\mu=(k^2-1)/(k^2+1)$.
The Miura transformation to the ``standard'' discrete KdV equation was presented in [\scimi]:
$$x_{n,m}={(k^2-1)w_{n,m+1}+(k^2+1)w_{n+1,m}\over2kw_{n,m}}.\eqdef\dpen$$
Whenever $w_{n,m+1}$ and $w_{n+1,m}$ happen to be related by $\mu w_{n,m+1}+w_{n+1,m}=0$, a zero appears at position $(n+1,m+1)$. (If the relation is $ w_{n,m+1}+\mu w_{n+1,m}=0$ then an infinity appears at that point). This zero is isolated, meaning that the singularity is confined then and there. Of course, more complicated situations may exist, when the relation on a diagonal holds over more adjacent points. A simple example is given in Figure \figdef\eight. 
\medskip
\centerline{\includegraphics[width=6 cm,keepaspectratio]{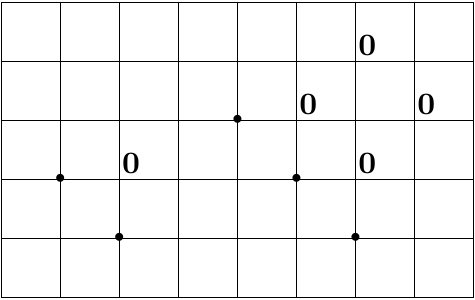}}
\smallskip{\bf Figure \eight}. {\sl  Singularities of the mKdV lattice, eq. (\dtri). The bullets indicate the points at which the values of $w$ must satisfy a relation, for the 0 to appear}
\smallskip
Given the form of the equation, we can, by simple inspection, conclude that if $w_{n,m}=0$, then $w_{n\pm1,m\pm1}=0$, populating an infinite oblique line with zeros. (And the same is true if $w_{n,m}=\infty$).

Guessing the taishi singularity is not straightforward and so we shall rely on the Miura transformation relating the solutions of (\dtri) to those of the Levi-Yamilov m-KdV. From (\dtes) we have
$${w_{n+1,m+1}\over w_{n,m}}={k+v_{n,m}\over k-v_{n,m}}\eqdaf\dhex$$
$${w_{n,m+1}\over w_{n+1,m}}={1-kv_{n,m}\over 1+kv_{n,m}},\eqno(\dhex{\rm b})$$
and taking into account that the condition for the taishi of (\ztes) is $v_{n,m+1}v_{n,m}=1$ we find for the taishi of (\dtri):
$$w_{n,m+1}+w_{n,m-1}=0.\eqdef\dhep$$
It is easy to verify that if (\dhep) is valid at one point, it is valid for all $n$.

We now turn to the interaction of a taishi with an oblique line populated by zeros. (Similar results are obtained if we consider a line of infinities). The simplest case is the interaction of a taishi of weight 1 with a line equally of weight 1. (As we previously explained, it suffices to consider interactions with a line of weight 1, the outcome from an interaction with a line of higher weight being identical to the interaction with a proper number of lines of weight one). Figure \figdef\nine\ shows the result of such an interaction.
\medskip
\centerline{\includegraphics[width=4.5 cm,keepaspectratio]{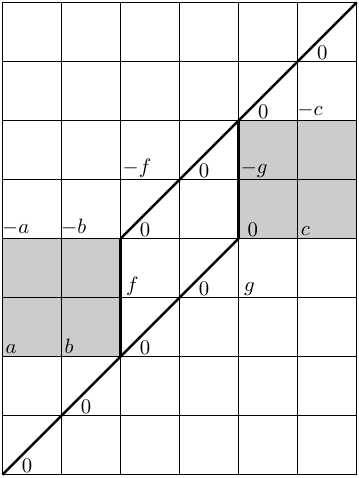}}
\smallskip{\bf Figure \nine}. {\sl  An interaction of a taishi of weight 1 with an infinite line also of formal weight 1. Notice the values $f,-f$ and $g,-g$ in the ``intermediate square'' during the interaction}
\smallskip
Already by looking at Figure \nine\ one can conclude that the dynamics of the taishi interaction in this case follow that of the ``standard'' KdV and not those of the Levi-Yamilov equations: the taishi is twice shifted upwards. As in the case of the KdV equation the effective weight of the oblique line is twice its formal weight. A further confirmation is obtained when one considers the interaction involving a taishi of weight 2, shown in Figure \figdef\ten. The taishi is now shifted upwards only once.
\medskip
\centerline{\includegraphics[width=4.5 cm,keepaspectratio]{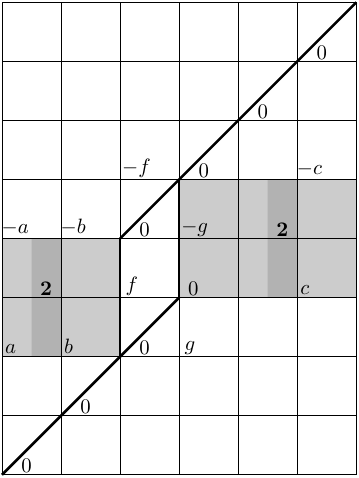}}
\smallskip{\bf Figure \ten}. {\sl  An interaction of a taishi of weight 2 with an infinite line of formal weight 1. The weight of the taishi is given before and after the interaction}
\smallskip

As in all previously studied cases the taishi of mKdV can also split and collapse following interaction. An example of the former is shown in Figure \figdef\eleven.
\medskip
\centerline{\includegraphics[width=5 cm,keepaspectratio]{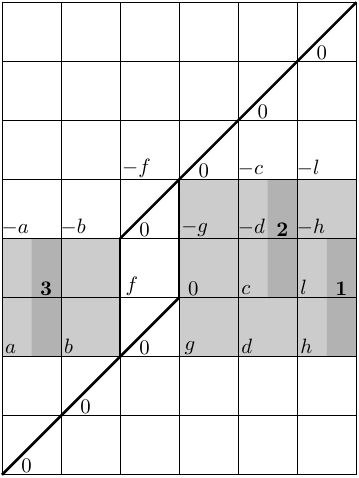}}
\smallskip{\bf Figure \eleven}. {\sl  An interaction of a taishi of weight 3 with an infinite line of formal weight 1. Two taishi emerge, partially overlapping, with weights 1 (lower) and 2 (higher)}
\medskip
An example of collapsing taishi is presented in Figure \figdef\twelve.
\medskip
\centerline{\includegraphics[width=6 cm,keepaspectratio]{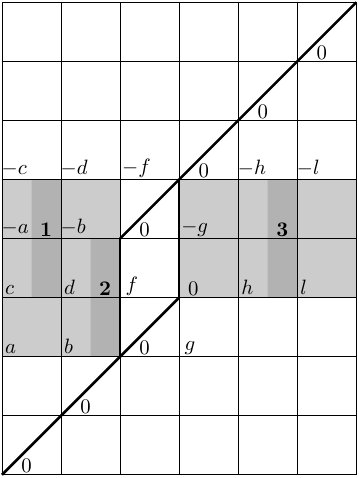}}
\smallskip{\bf Figure \twelve}. {\sl  An interaction of two taishi of weights 2 and 1 with an infinite line of formal  weight 1. A single taishi emerges, with weight 3}
\smallskip
At this point it is interesting, in view of the Miura relations (\dtes), to look for the simultaneous behaviour of the taishi of standard mKdV, (\dtri) and the Levi-Yamilov (\ztes) one, under interaction. The interaction of a taishi of (\dtri) with an oblique line of zeros with weight 1 was shown in Figure \nine. In Figure \figdef\thirteen\ we show the same interaction, adding the evolution of the Levi-Yamilov taishi. 
\medskip
\centerline{\includegraphics[width=9 cm,keepaspectratio]{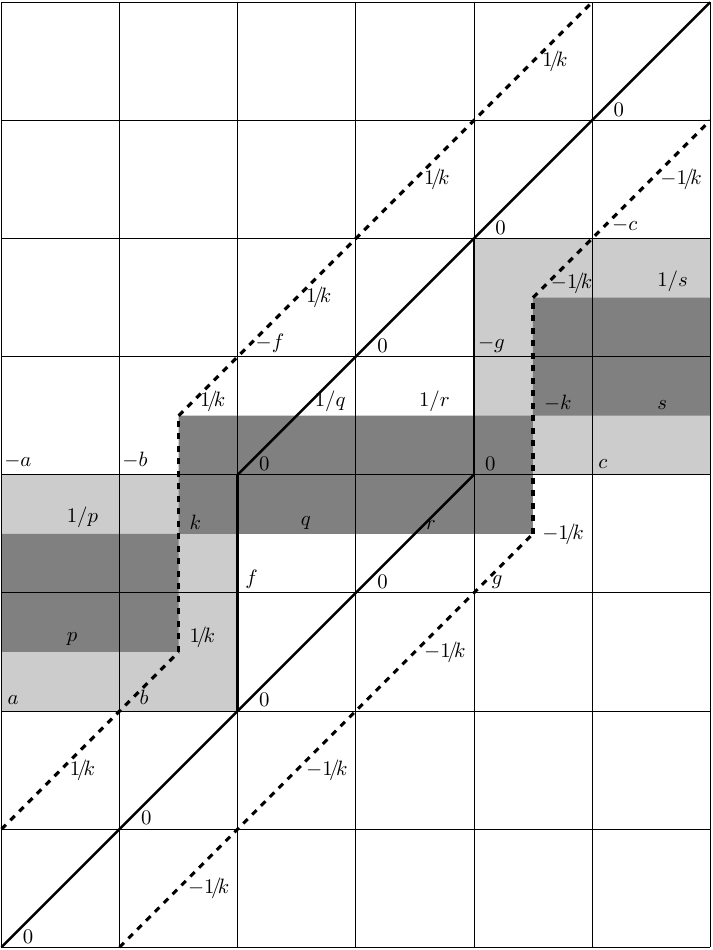}}
\smallskip{\bf Figure \thirteen}. {\sl  A weight 1 taishi of eq. (\dtri) interacting with an infinite line of weight 1, shown in light grey. (It is exactly the same as the one shown in Figure \nine). The taishi of eq. (\ztes), obtained through the Miura (\dtes), is shown in dark grey}
\smallskip
We remark that the taishi of (\ztes) undergoes two interactions, with, as a result, a double shift upwards. The reason for this is the existence of not just one infinite line of $1/k$, as in all cases studied in [\ourmkdv], but of a second one, with values $-1/k$, symmetric with respect to the infinite line of zeros of the mKdV (\dtri). The existence of this line is a direct consequence of the Miura transformation (\dtes). Whenever $w_{n+l,m+l}=0$ for all $l$, we have $v_{n+l,m+1+l}=1/k$ {\sl and} $v_{n+l,m-1+l}=-1/k$. Thus the taishi in terms of $v$ encounters two oblique lines and undergoes two interactions. 

As we explained above, the taishi interaction follows the same symbolic dynamics as that of the ``standard'' KdV, which means that every interaction of a taishi with an oblique line of formal weight 1 necessitates two steps. As a consequence the interactions, depicted in figures \ten, \eleven\ and \twelve, can be represented by the diagrams below:
$$\matrix{ 0\cr 0\cr 0\cr \bf2\cr 0\cr 0}\Longrightarrow\matrix{0\cr0\cr 1\cr1\cr0\cr0}\Longrightarrow\matrix{0\cr0\cr \bf2\cr0\cr0\cr0}\hskip2.5cm\matrix{ 0\cr 0\cr0\cr \bf3\cr 0\cr 0}\Longrightarrow\matrix{0\cr 0\cr1\cr2\cr0\cr0}\Longrightarrow\matrix{0\cr 0\cr\bf2\cr\bf1\cr0\cr0}\hskip2.5cm\matrix{ 0\cr 0\cr\bf1\cr \bf2\cr 0\cr 0}\Longrightarrow\matrix{0\cr0\cr2\cr1\cr0\cr0}\Longrightarrow\matrix{0\cr 0\cr\bf3\cr0\cr0\cr0}$$
The numbers in boldface correspond to the weights in the taishi before interaction (on the left) and after it has left the interaction region (on the right).
\section{A sine-Gordon equation}
The third lattice system we are going to study is the sine-Gordon equation proposed by Heredero and collaborators [\heredero]. It has the form
$$y_{n+1,m+1}y_{n,m}=\left({\mu +y_{n+1,m}\over 1+\mu y_{n+1,m}}\right)\left({\mu +y_{n,m+1}\over 1+\mu y_{n,m+1}}\right),\eqdef\doct$$
where $\mu$ is a parameter. This is  not the ``standard'' form of the lattice sine-Gordon equation. The latter, derived by Hirota [\refdef\sinegordo], is
$$z_{n+1,m+1}z_{n,m}={\mu +z_{n+1,m}z_{n,m+1}\over 1+\mu z_{n+1,m}z_{n,m+1}},\eqdef\denn$$
and the two equations are related by the Miura
$$y_{n,m}=z_{n+1,m}z_{n,m+1}.\eqdef\ddek$$
Going back to equation (\doct), we remark readily that if $y_{n,m}+\mu=0$, then we have $y_{n+1,m}=0$ and $y_{n,m+1}=0$. (Similarly if $1+\mu y_{n,m}=0$ we have $y_{n+1,m}=\infty$ and $y_{n,m+1}=\infty$). However these zeros (or infinities) are isolated singularities, meaning that iterating further (\doct) leads to finite values. The structure of the singularities does not become more complicated when we consider the relation $y+\mu=0$ to be valid on two adjacent points $(n+1,m)$ and $(n,m+1)$: the zeros appearing at $(n+2,m)$, $(n+1,m+1)$ and $(n,m+2)$ are immediately confined. On the other hand if we have adjacent values of $-\mu$ and $-1/\mu$ the singularity pattern becomes more complicated. An example is given in Figure \figdef\fourteen .

An oblique line of singularities, running in the SW-NE direction, also exists. It consists of alternating zeros and infinities. 
\medskip
\centerline{\includegraphics[width=8 cm,keepaspectratio]{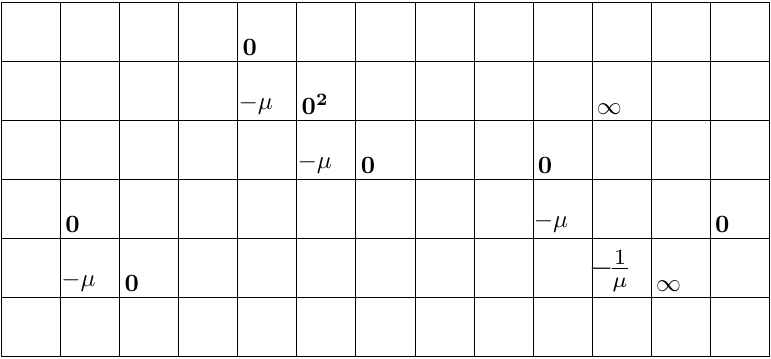}}
\smallskip{\bf Figure \fourteen}. {\sl  Confined singularities of equation (\doct). Here $0^2$ means that the weight of the singular value 0 is equal to 2}
\smallskip

The condition for the existence of a taishi type singularity in the case of (\doct) is more complicated.. We shall not go into the details of its derivation. Suffice it to say that the latter is based on the well-known relation between m-KdV and sine-Gordon, studied in detail in [\scimi]. It turns out that if the relation
$$y_{n,m}+y_{n,m+1}+\mu(1+y_{n,m}y_{n,m+1})=0\eqdef\vena$$
is satisfied for some $n$, it is valid for all $n$.

Figure \figdef\fifteen\ gives an example of an interaction of a taishi with an oblique line of singularities. If the interaction is such that the vertical jump of the oblique line is from infinity to infinity, then the value appearing on the vertical between the two infinities is $-1/\mu$, instead of the value $-\mu$ which appears when the jump is between two zeros, as in the example above. 
\medskip
\centerline{\includegraphics[width=6 cm,keepaspectratio]{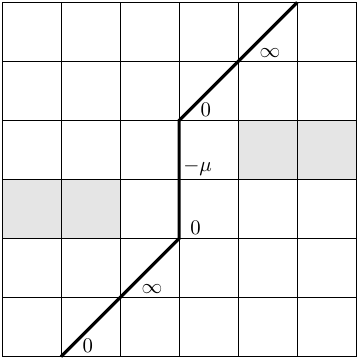}}
\smallskip{\bf Figure \fifteen}. {\sl  The interaction of a taishi of (\doct) with an oblique line of singularities}
\smallskip

We remark that the symbolic dynamics of the interaction follow the prescription presented in the introduction for the Levi-Yamilov m-KdV equation, and in section 3 for the Levi-Yamilov KdV.
\section{Discussion}
In this work we have examined three integrable lattice equations from the point of view of the existence of singularities with an emphasis on their interactions. In all three cases we encountered the same behaviour already noticed in our previous studies [\kdv] and [\ourmkdv]. The typology of the singularities can be simply summarised as follows.

Three distinct types of singularities exist. The first corresponds to singularities that have finite extent: once they appear they disappear after a finite number of steps. The patterns of said singularities, on the plane, do vary with the equation but their common feature is that they are confined. The second type of singularities are of infinite extent, populating a line that runs in a direction that can be chosen as south-west to north-east, with fixed values. The values that appear on this infinite line depend on the equation at hand. The third type of singularity populate a horizontal (or vertical) strip, a structure we have dubbed `taishi'. The conditions for the existence of a taishi do vary with the equation and can, in some cases, be somewhat involved. 

The singularities interact with each other. In particular the taishi exhibits a great variety of behaviour when interacting with an infinite oblique line of singularities. Despite the complicated aspect of these interactions the underlying process is simple and can be formulated in terms of symbolic dynamics, consisting in a small collection of simple rules.  Typically, for horizontal taishis interacting with a singularity on a diagonal in the plane, one moves upwards, subtracts 1 from the weight of the first encountered non-zero strip and adds 1 to the strip just above and proceeds in the same way up to the last existing non-zero weight strip. Moreover this prescription is essentially the same for all equations studied (but for some of them it has to be applied twice while for some other a single application suffices). 

Two interesting remarks are in order at this point. The symbolic dynamics we just referred to are precisely the dynamics of a Box \& Ball system with carrier. The boxes have unlimited capacity, the weights of the taishi play the role of the balls and the carrier has a capacity of 1. An empty carrier does nothing when it encounters an empty box, but loads one ball if the box is not empty and then unloads it at the next box. It is remarkable that the Box \& Ball  system just described can be obtained from the ultradiscrete limit of the modified KdV lattice equation [\takama]. The latter leads to a cellular automaton describing the interaction of the solitons of the equation but finding that it describes also the interaction of the singularities is, to say the least, intriguing.

The second remark has to do with the fact that all the equations examined here and in the previous publications (as well as those that we studied but opted not to present here so as not to overburden the paper with, after all, similar results) are equations that can be obtained by some proper reduction of the Hirota-Miwa equation [\refdef\dagte, \refdef\miwa].  So the question arises whether the properties studied here can be transposed to the case of the Hirota-Miwa three-dimensional lattice. Since the Hirota-Miwa equation has a most simple expression in terms of tau functions it would perhaps be preferable to recast the conclusions reached here in the bilinear formalism. In this way it might be possible to explain the emergence and universality of the symbolic dynamics governing the interaction of the singularities (but we are aware that this is a very tall order indeed).
 \paragraph{Acknowledgments.}
 RW  would like to thank the Japan Society for the Promotion of Science (JSPS) for financial support through the KAKENHI grant 23K22401. 
%
%

\begin{description}
\item{[\bountis]} A. Ramani, B. Grammaticos and T. Bountis, The Painlev\'e property and singularity analysis of integrable and non-integrable systems, Physics Reports 180 (1989) 159-245.
\item{[\sincon]} B. Grammaticos, A. Ramani and V. Papageorgiou, Do integrable mappings have the Painlev\'e, property?, Phys. Rev. Lett. 67 (1991) 1825-1828.
\item{[\desoto]} B. Grammaticos, R. Willox, T. Mase and M. Kanki, The redemption of singularity confinement, J. Phys. A 48 (2015) 11FT02.
\item{[\hirota]} R. Hirota,  Nonlinear Partial Difference Equations I: A Difference Analogue of the Korteweg-deVries Equation, J. Phys. Soc. Japan 43 (1977) 1424-1433.
\item{[\wynn]} P. Wynn, Singular rules for certain non-linear algorithms, BIT 3 (1963), 175-195.
\item{[\kdv]} D. Um, A. Ramani, B. Grammaticos, R. Willox and J. Satsuma, On the singularities of the discrete Korteweg-deVries equation, J. Phys. 54 (2021) 095201.
\item{[\levi]}  D. Levi and R.I. Yamilov, On a nonlinear integrable difference equation on the square, Ufimsk. Mat. Zh. 1 (2009) 101-105.
\item{[\takasa]} D. Takahashi and J. Satsuma, A soliton cellular automaton, J. Phys. Soc. Japan 59 (1990) 3514-3519.
\item{[\takama]} D. Takahashi and J. Matsukidaira, Box and Ball system with a carrier and ultradiscrete modified KdV equation, J. Phys. A: Math. Gen. 30 (1997) L733.
\item{[\capel]} H. Capel, F.W. Nijhoff and V. Papageorgiou, Complete integrability of Lagrangian mappings and lattices of KdV type, Phys. Lett A 153 (1991) 377-387.
\item{[\heredero]} R.H. Heredero, D Levi, M. Petrera and C. Scimiterna, Multiscale expansion on the lattice and integrability of partial difference equations, J. Phys. A 41 (2008) 315208.
\item{[\notso]} A. Ramani, B. Grammaticos, R. Willox and J. Satsuma, On two (not so) new integrable partial difference equations, A. Ramani, B. Grammaticos, J. Phys. A 42 (2009) FT 282002.
\item{[\ourmkdv]} B. Grammaticos, T. Tamizhmani and R. Willox, On the singularity structure of a discrete modified-Korteweg-deVries equation, J. Phys. A 55 (2022) 265203.
\item{[\scimi]} B. Grammaticos, A. Ramani, C. Scimiterna and R. Willox,  Miura transformations and the various guises of integrable lattice equations, J. Phys. A 44 (2011) 152004
\item{[\hirom]} R. Hirota, Discretization of the Potential Modified KdV Equation, J. Phys. Soc. Japan 672 (1998) 2234-2236.
\item{[\sinegordo]} R. Hirota, Nonlinear Partial Difference Equations III: Discrete Sine-Gordon Equation, J. Phys. Soc. Japan 43 (1977) 2079-2086.
\item{[\dagte]} R. Hirota, Discrete Analogue of a Generalized Toda Equation, J. Phys. Soc. Japan 503 (1981) 3785-3791.
\item{[\miwa]} T. Miwa, On Hirota's difference equations, Proc. Japan Acad. 58 (1982) 9-12.
\end{description}
\end{document}